\documentstyle[aps]{revtex}

\newcommand{\BE}{\begin{equation}}
\newcommand{\EE}{\end{equation}}
\newcommand{\BA}{\begin{eqnarray}}
\newcommand{\EA}{\end{eqnarray}}

\begin{document}
\draft
%\tightenlines

\twocolumn[\hsize\textwidth\columnwidth\hsize\csname@twocolumnfalse\endcsname
\title{Comment on ``Multiple scattering in a reflecting cavity: Application to fish counting in a tank
[J. Acoust. Soc. Am. {\bf 109}, 2587-2597 (2001)]" }
\author{Zhen Ye$^1$ and Dezhang Chu$^2$}
\address{$^1$Wave Phenomena Laboratory, Department of Physics, National Central University, Chung-li,
Taiwan 32054 \\
$^2$Department of Applied Ocean Physics and Engineering, Woods
Hole Oceanographic Institute, Woods Hole, MA 02543}

\date{\today}

\maketitle

\begin{abstract}
This paper presents a comment on the recent work on fish counting
in a tank (J. Acoust. Soc. Am. {\bf 109}, 2587-2597 (2001)). It is
pointed out that there are ambiguities with the counting method.
\end{abstract}

 ]

%\twocolumn
%\narrowtext

%\newpage

In a recent paper\cite{Roux}, de Rosny and Roux proposed a method
for counting fish in a tank, an important issue for the fisheries
community. As far as we know, this is the first published report
on counting fish in a tank. We recognize that the fish information
may have been detected by the method. However, in its present form
of data analysis, we believe that the foundations for the analysis
are questionable. As a matter of fact, their approach has failed
to provide reasonable scattering properties of fish. The reasons
are elaborated below.

(1) The multiple scattering in a reflecting cavity. The basis for
the fish counting method in \cite{Roux} is described by Fig.~1 of
\cite{Roux}. The analysis relies on the treatment of the multiple
scattering picture in this figure. In the context, the authors
argued that the effect of the reflecting boundaries of a tank can
be regarded as mirrors, and thus acoustic scattering in such a
system is regarded as a medium of scatterers and their images
without the boundaries. The multiple reflections make the system
equivalent to a system of infinite size.

The first ambiguity in the later development is that effects from
the images of the acoustic sources are not considered in
\cite{Roux}. Although the images of the sources are shown in
Fig.~1, their possible effects are not discussed anywhere in
\cite{Roux}. In addition, the reflection from the water surface is
not taken into account either. The presence of the water surface
also acts as a mirror, giving rise to interactions between the
scatterers and their images due to this mirror. It is worth
pointing out that the water surface acts as a pressure release
plane, which has different effects from the presumably
total-reflecting side boundaries of the tank.

The second ambiguity lies in the way how the multiple scattering
among the scatterers and the images in Fig.~1 is accounted for, a
crucial problem in the analysis. In \cite{Roux}, the authors wrote
down two formulas in Eqs.~(1) and (2) for total and coherent
intensity as a result of the multiple scattering; here we note
that in \cite{Roux} Eqs.~(1) and (2) have been mistakenly regarded
as the coherent intensity and incoherent intensity respectively -
the detailed information about the definitions of these quantities
can be referred to \cite{Ishimaru}. With reference to
\cite{Ishimaru}, it is ready to verify that both Eqs.~(1) and (2)
are actually the results of the multiple scattering when wave
propagates in a cloud of {\it completely independent} random
scatterers (Refer to, e.~g. Chapter 14 in \cite{Ishimaru}). In the
system such as many scatterers in a reflecting cavity, the
scatterers and their images are {\it not} independent scatterers.
Thus Eqs.~(1) and (2) are not applicable to the present case.
There is a rich body of literature on acoustic scattering by
bodies in the presence of boundaries (e.~g.
\cite{Young1,Smith,Faw,Tolstoy}). Correlated scatterers can give
rise to interference between multiply scattered waves that is
absent in a group of independent scatterers. Take the simplest
case of an object near a boundary as example. The multiple
scattering from a scatterer and its image differs from the
scattering from two actual scatterers. This can be easily inferred
from, for example, Ref.~\cite{Tolstoy}.

(2) The mathematical derivation. The results of Ref.~\cite{Roux}
rely mainly on its Eq.~(4). In writing down this equation, the
authors made the assertion ``Eq.~(3) can be written (as Eq.~(4))
by changing from a space variable to a time variable", then
Eq.~(4) is obtained. But no details were given. The derivation
from Eq.~(3) to Eq.~(4) is mathematically unclear. Eq.~(3) denotes
how wave is attenuated after travelling a distance, but Eq.~(4)
represents the time evolution of backscattered signals. The former
is a {\it forward} propagating process, while the later is a {\it
backscattering} process. The link between the two is not obvious.
In fact, the time evolution of backscattered signals has been
detailed in Chapter 5 of \cite{Ishimaru}. It is not evident how
the time series of the ratio between the coherent and the total
backscattered signals can be represented by a simple equation like
Eq.~(4).

(3) The experimental results. The method of \cite{Roux} yields
doubtful scattering properties of fish. The main goal of
\cite{Roux} is to obtain accurate total scattering cross section
of fish. Two fish species are measured: zebra fish with length
about 1 cm at 400 kHz and 35 cm long striped bass at 12.8 kHz.
Take the bass fish as the example. According to \cite{Roux}, the
total scattering cross section of the strip bass is $4\pi R^2$
with $R \sim 3.8$ cm, leading to $\sigma_s \sim 181$ cm$^2$. This
value is much larger than what would be expected for a fish of 35
cm length. The reasons follow.

First, the total scattering cross section is a computable
quantity. According to the optical theorem\cite{Ishimaru}, the
total cross section can be calculated from the imaginary part of
the forward scattering function of a scatterer \BE \sigma_t =
\sigma_s + \sigma_a = \frac{4\pi}{k} \mbox{Im}[f(0)],\EE where
$f(0)$ is the scattering function in the forward scattering
direction, $\sigma_s$ and $\sigma_a$ are the total scattering
cross section and the absorption cross section respectively; the
absorption can be caused by such effects as thermal exchange or
viscosity of the scatterer. For frequencies above the resonance of
the fish swimbladder like the cases considered in \cite{Roux}, the
absorption due to the swimbladder is negligible. Furthermore the
acoustic absorption due to fish bodies is also negligible.
Therefore we have \BE\sigma_s \approx \sigma_t =\frac{4\pi}{k}
\mbox{Im}[f(0)]. \EE This applies to scattering by both fish
swimbladders and fish bodies. The advances in modelling acoustic
scattering by fish allow for reasonable estimates of fish
scattering cross sections. Fish scattering models can be referred
to, e.~g., the textbook\cite{Clay}. Indeed, the optical theorem
has been applied to existing experimental measurements of various
fish scattering cross sections, yielding encouraging
agreements\cite{fish}. Applying the method in \cite{fish}, we
estimate that the total scattering cross section for a 35 cm long
fish at 12.8 kHz is around 30 - 40 cm$^2$, depending on the aspect
ratio used in the modelling. This is much smaller than the value
obtained in \cite{Roux}. It is our experience that the scattering
properties vary only slightly for different fish species with the
same length.

Second, there exist experimental data on acoustic scattering on
fish of similar length. For example, the previous experiment data
on Saithe fish of 35.1 cm length yields a total scattering cross
section around 30 cm$^2$\cite{Foote}, in the same order as the
theoretical value\cite{fish}. This experiment was carried out at a
frequency higher than 12.8 kHz used in \cite{Roux}. From the
modelling, this would lead to a slightly higher value in the total
scattering cross section than that at 12.8 kHz. It is therefore
reasonable to conclude that the total scattering cross section for
a 35 cm long fish at 12.8 kHz would be in the order of a few tens
of square centimeters, which is considerably smaller than that
obtained in \cite{Roux}.

In conclusion, the method in \cite{Roux} requires further
development and elaboration in order to be applicable to counting
fish in an aquatic tank.

{\bf Acknowledgments} Dr. P.-G. Luan is thanked for bringing
Ref.~\cite{Roux} to our attention. We thank Dr. Tim Stanton (WHOI)
for useful correspondence. The work received support from the
National Science Council through a grant to ZY.

%\section*{References}


\begin{references}

\bibitem{Roux} J. de Rosny and P. Roux, ``Multiple scattering in a reflecting cavity:
Application to fish counting in a tank", J. Acoust. Soc. Am. {\bf
109}, 2587-2597 (2001).

\bibitem{Ishimaru} A. Ishimaru, {\it Wave propagation and scattering in
random media}, (Academic Press, New York, 1978), Vols. I and II.

\bibitem{Young1} J. C. Bertrand and J. W. Young, ``Scattering
between a cylinder and a plane", J. Acoust. Soc. Am. {\bf 60},
1265-1269 (1976).

\bibitem{Smith} G. C. Bishop and J. Smith, ``Scattering from an elastic shell and
a rough fluid-elastic interface", J. Acoust. Soc. Am. {\bf 101},
767-788 (1997)

\bibitem{Faw} J. A. Fawcett, W. L. J. Fox, and A. Maguer, J.
Acoust. Soc. Am. {\bf 104}, 3296-3304 (1998).

\bibitem{Tolstoy} I. Tolstoy, ``Superresonant systems of
scatterers. I", J. Acoust. Soc. Am. {\bf 80}, 282-294 (1986).

%\bibitem{YF} Z. Ye and C. Feuillade, ``Sound scattering by an air
%bubble near a plane sea surface", J. Acoust. Soc. Am. {\bf 102},
%798-805 (1997).

\bibitem{Clay} H. Medwin and C. S. Clay, {\it Acoustical
Oceanography}, (Academic Press, New York, 1997).

\bibitem{fish} Z. Ye, ``On acoustic attenuation by swimblader fish",
J. Acoust. Soc. Am. {\bf 100}, 669-672 (1996)

\bibitem{Foote} K. G. Foote, ``Analysis of empirical observations
on the scattering of sound by encaged aggregations of fish",
Fisheridir. Skr. Ser. Havunders. {\bf 16}, 422-455 (1978).
\end{references}
\end{document}